\documentstyle[art14]{article}
\def\t2{\tilde t}
\def\u2{\tilde u}
\def\s2{\tilde s}
\begin{document}
\begin{center}
 {\bf INTRINSIC CHARM CONTRIBUTION TO $J/\psi$ PLUS OPEN CHARM
  ASSOCIATED PRODUCTION AT HIGH ENERGIES}

A.P.Martynenko and V.A.Saleev

Samara State University, Samara, 443011, Russia

\end{center}

\begin{abstract}
 We present the results of a calculation of the intrinsic charm
 quark contribution to $J/\psi$ plus open charm associated production
 at large transverse
 momentum in hadron collisions at high energies. It is shown that the
 value  of the cross section for such process and its energy dependence
 significantly determine by the choice of  charm quark distribution
 function.   We find that
 intrinsic charm contribution to inclusive $J/\psi$ hadroproduction
 is about 1\% of the total contribution of all mechanisms.
\end{abstract}

The study of charmonium production at hadron colliders is presently an
active
field of theoretical \cite{1,2,3} and experimental researches
\cite{4,5,6}.
Previous calculations of $J/\psi$ production at large $p_\perp$ in $pp$
and $p\bar p$ collisions have included direct charmonium production via
gluon-gluon fusion \cite{1,7} and production via decay $B\to J/\psi X$
\cite{2}.
Recently it was presented the preliminary results of a calculation of
the
gluon and charm quark fragmentation contribution to $J/\psi$ production
 at
large $p_\perp$ \cite{8}. Here we examine the intrinsic charm
contribution
to $J/\psi$ hadroproduction. Based on perturbative QCD and
nonrelativistic
quark model, our calculation  is the same in part the approach used
in \cite{9} for description of pion electroproduction on protons.

The calculation of the intrinsic charm quark contribution to associated
$J/\psi$ plus open charm hadroproduction and photoproduction \cite{10}
 is very important for study of the charm quark distribution function
 in
 a proton \cite{11}. The study of inclusive $J/\psi$ production is of
 particular
 importance because the decay $B\to J/\psi X$ plays a crucial role in
 the
 measurement of $b$ quark production. In order to understand this
 phenomenon, it is necessary to take into consideration the background
 due to direct production of $J/\psi$.

We start from amplitudes, which correspond  diagrams in  Fig.1.
As usual in potential model, the $J/\psi$ is represented as a
nonrelativistic quark-antiquark bound system in singlet colour state
with specified mass $M=2m$ ($m$ is charm quark mass) and spin-parity
$J^p=1^-$.  The amplitudes of the $J/\psi$ production have
 the following forms:
\begin{equation}
 M_1=g^3T^cT^aT^a\bar U(q')\hat\varepsilon_{g}\frac{\hat q'-\hat k+m}
 {(k-q')^2-m^2}\gamma^{\mu}\frac{\hat P}{(p-q)^2}
  \gamma_{\mu}U(q),
\end{equation}
\begin{equation}
M_2=g^3T^aT^aT^c \bar U(q')\gamma^{\mu}\frac{\hat P}{(p+q')^2}
\gamma_{\mu}
\frac{\hat k+\hat q+m}{(k+q)^2-m^2}\hat\varepsilon_{g}U(q),
\end{equation}
\begin{equation}
M_3=g^3T^aT^cT^a\bar U(q')\gamma^{\mu}\frac{\hat P}{(q'+p)^2}\hat
\varepsilon_g
\frac{\hat p-\hat k+m}{(p-k)^2-m^2}\gamma_{\mu}U(q),
\end{equation}
\begin{equation}
M_4=g^3T^aT^cT^a\bar U(q')\gamma^{\mu}
\frac{\hat k-\hat p+m} {(p-k)^2-m^2}\varepsilon_{g}
\frac{\hat P}{(p-q)^2}\gamma_{\mu}U(q),
\end{equation}
\begin{equation}
 M_5=-ig^3f^{bac}T^bT^a\bar U(q')\gamma^{\rho}\frac{\hat P}{(p+q')^2}
 \frac{\gamma^{\sigma}}{(q-p)^2}U(q)C_{\mu\sigma\rho}\varepsilon^
  {\mu}_g,
\end{equation}
where
$$C_{\mu\sigma\rho}=(q+q')_{\mu}g_{\sigma\rho}-(p+k+q')_{\sigma}
  g_{\mu\rho}+(k-q+p)_{\rho}g_{\mu\sigma}.$$
In these formulas:
$$\hat P=\frac{F_c}{\sqrt 2}A\hat\varepsilon_J(\hat p+m),$$
$A=\Psi(0)/\sqrt m$,  $F_c=\delta^{kr}/\sqrt{3}$,  k and  r  are  colour
indexes of charm quarks,
$T^a=\lambda^a/2$.  It is well known that $\Psi(0)$, which is
equal to $J/\psi$ wave function at the origin,  can be extracted  in
the lowest  order  of perturbative QCD from the leptonic decay
width of the $J/\psi$:
\begin{equation}
  \Gamma_{ee}=4\pi e_q^2\alpha^2\frac{|\Psi(0)|^2}{m^2}.
\end{equation}
We shall put in our calculation $\Gamma_{ee}=5.4$ KeV \cite{12} and $m=
1.55$ GeV.

If we average and sum over spins and  colours  of  initial  and  final
particles, we obtain the expression for square of matrix element:
  \begin{equation}
  |\bar M|^2=B_{gc}\sum_{i\le j=1}^{5}C_{ij}K_{ij}(\s2,\t2,\u2),
   \end{equation}
    where
  $$B_{gc}=\frac{\pi^2\alpha_s^3\Gamma_{ee}}
    {16\alpha^2m},$$
$$\tilde s=\hat s/m^2,\qquad \tilde t=\hat t/m^2,\qquad \tilde  u=\hat
u/m^2,$$
$\hat s, \hat t, \hat u$ are usual Mandelstam variables
and $\hat s+\hat t+\hat u=6m^2$. The explicit analytical formulae for
functions $K_{ij}$ are:
\begin{eqnarray}
      K_{11}&=&-4(2\s2\t2-2\s2+\t2^2\u2-4\t2^2-8\t2\u2+14
      \t2+7\u2-106)\nonumber\\
     &&/(\t2^4-4\t2^3+6\t2^2-4\t2+1)
\end{eqnarray}
\begin{eqnarray}
      K_{12}&=&4(\s2^3-\s2^2\t2-6\s2^2-\s2\t2^2-2\s2\t2\u2+16
      \s2\t2-\s2\u2^2+8\s2\u2\nonumber\\
      && -28\s2+\t2^3-6\t2^2-\t2\u2
      ^2+8\t2\u2-28\t2+4\u2^2-126\u2+276)/\nonumber\\
      && (\s2^2\t2^2-2\s2^2\t2+\s2^2-2\s2\t2^2+4\s2\t2-
      2\s2+\t2^2-2\t2+1)
\end{eqnarray}
\begin{eqnarray}
      K_{13}&=&8(\s2^2\t2+\s2^2+2\s2\t2^2+\s2\t2\u2-22\s2\t2+7
      \s2\u2+16\s2+\nonumber\\
&&  2\t2^2\u2-24\t2^2-11\t2\u2+217
 \t2-2\u2^2+41\u2-511)/(\s2\t2^2\u2-4\s2\nonumber\\
&& \t2^2-2\s2\t2\u2+8\s2\t2+\s2\u2-4\s2-\t2^2\u2+4
      \t2^2+2\t2\u2-8\t2-\u2+4)
\end{eqnarray}
\begin{eqnarray}
      K_{14}&=&-8(\s2^3+\s2^2\t2-18\s2^2+\s2\t2^2+\s2\t2\u2-16
      \s2\t2-\s2\u2^2-3\s2\u2+\nonumber\\
       &&166\s2-11\t2^2-5\t2\u2+
 115\t2+13\u2^2-65\u2-335)/\nonumber\\
 &&(\t2^3\u2-4
      \t2^3-3\t2^2\u2+12\t2^2+3\t2\u2-12\t2-\u2+4)
\end{eqnarray}
\begin{eqnarray}
      K_{22}&=&-4(\s2^2\u2-4\s2^2+2\s2\t2-8\s2\u2+14\s2-2
      \t2+7\u2-106)/\nonumber\\
      &&(\s2^4-4\s2^3+6\s2^2-4\s2+1)
\end{eqnarray}
\begin{eqnarray}
      K_{23}&=&-4(\s2^2\t2-11\s2^2+\s2\t2^2+\s2\t2\u2-16\s2\t2-
      5\s2\u2+115\s2+\nonumber\\
      &&\t2^3-18\t2^2-\t2\u2^2-3\t2\u2+
      166\t2+13\u2^2-65\u2-335)/\nonumber\\
        &&(\s2^3\u2-4
      \s2^3-3\s2^2\u2+12\s2^2+3\s2\u2-12\s2-\u2+4)
\end{eqnarray}
\begin{eqnarray}
      K_{24}&=&8(2\s2^2\t2+2\s2^2\u2-24\s2^2+\s2\t2^2+\s2\t2
      \u2-22\s2\t2-11\s2\u2+\nonumber\\
     &&  217\s2+\t2^2+7\t2\u2+16
      \t2-2\u2^2+41\u2-511)/(\s2^2\t2\u2-4\s2^2\t2\nonumber\\
     && -\s2^2\u2+4\s2^2-2\s2\t2\u2+8\s2\t2+2\s2
      \u2-8\s2+\t2\u2-4\t2-\u2+4)
\end{eqnarray}
\begin{eqnarray}
      K_{33}&=&-16(2\s2^2+2\s2\t2+6\s2\u2-46\s2+\t2^2\u2-10
      \t2^2-\nonumber\\
       &&4\t2\u2+74\t2+2\u2^3-22\u2^2+53\u2-118)/
 (\s2^2\u2^2-8\s2^2\u2+\nonumber\\
   && 16\s2^2-2\s2\u2
      ^2+16\s2\u2-32\s2+\u2^2-8\u2+16)
\end{eqnarray}
\begin{eqnarray}
      K_{34}&=&-16(2(\s2^2+\s2\t2-11\s2+\t2^2+2\t2\u2-19\t2+\u2
      ^3-9\u2^2+28\u2+11))/\nonumber\\
       &&(\s2\t2\u2^2-8\s2
      \t2\u2+16\s2\t2-\s2\u2^2+8\s2\u2-16\s2\nonumber\\
       &&-\t2\u2^2+8
      \t2\u2-16\t2+\u2^2-8\u2+16)
\end{eqnarray}
\begin{eqnarray}
      K_{44}&=&-16(\s2^2\u2-10\s2^2+2\s2\t2-4\s2\u2+74\s2+2
      \t2^2+6\t2\u2-\nonumber\\
       &&46\t2+2\u2^3-22\u2^2+53\u2-118
      )/(\t2^2\u2^2-8\t2^2\u2+16\t2^2-2\t2\u2^2\nonumber\\
&& +16\t2\u2-32\t2+\u2^2-8\u2+16)
\end{eqnarray}
\begin{eqnarray}
      K_{55}&=&4(\s2^3-\s2^2\t2+\s2^2\u2-14\s2^2-\s2\t2
     ^2+2\s2\t2\u2\nonumber\\
&& +20\s2\t2-\s2\u2^2 -8\s2\u2+40\s2+
      \t2^3+\t2^2\u2-14\t2^2\nonumber\\
&& -\t2\u2^2-8\t2\u2+40\t2+2
     \u2^3+4\u2^2-4\u2-168)/\nonumber\\
&&(\s2^2\t2^2-2\s2^2\t2+\s2^2-2\s2\t2^
     2+4\s2\t2-2\s2+\t2^2-2\t2+1)
\end{eqnarray}
\begin{eqnarray}
      K_{51}&=&4(s2^3+\s2^2\t2-20\s2^2-\s2\t2^2+2\s2
      \t2\u2-12\s2\t2-\s2\u2^2\nonumber\\
&&       -4\s2\u2+168\s2-\t2^3-2 \t2^2\u2+24\t2^2+\t2\u2^2
          -12\t2\u2-88\t2+14\u2^2-232)\nonumber\\
&&      /(\s2\t2^3-3\s2\t2^2+3\s2\t2-\s2-\t2^3+
      3\t2^2-3\t2+1)
\end{eqnarray}
\begin{eqnarray}
      K_{52}&=&-4(\s2^3+\s2^2\t2+2\s2^2\u2-24\s2^2-\s2
      \t2^2-2\s2\t2\u2+12\s2\t2\nonumber\\
&&       -\s2\u2^2+12\s2\u2+88
     \s2-\t2^3+20\t2^2+\t2\u2^2+4\t2\u2-168\t2-14\u2^2
      +232)\nonumber\\
&&      /(\s2^3\t2-\s2^3-3\s2^2\t2+3\s2^2+3
      \s2\t2-3\s2-\t2+1)
\end{eqnarray}
\begin{eqnarray}
      K_{53}&=&8(4\s2^2-\s2\t2^2+22\s2\t2+10\s2\u2-
      117\s2+\t2^3+2\t2^2\u2\nonumber\\
&&      -26\t2^2-\t2\u2^2+2\t2\u2
      +69\t2+4\u2^3-29\u2^2+34\u2+80)\nonumber\\
&&  /(\s2^2\t2\u2-4\s2^2\t2-\s2^2\u2+4\s2^2 -2\s2\t2\u2\nonumber\\
&&   +8\s2\t2+2\s2\u2-8\s2+\t2\u2-4\t2-\u2+4)
\end{eqnarray}
\begin{eqnarray}
       K_{54}&=&8(\s2^3-\s2^2\t2+2\s2^2\u2-26\s2^2+22
      \s2\t2-\s2\u2^2+2\s2\u2\nonumber\\
&&       +69\s2+4\t2^2+10\t2\u2-
      117\t2+4\u2^3-29\u2^2+34\u2+80)\nonumber\\
&&      /((\s2\t2^2\u2-4\s2\t2^2-2\s2\t2\u2+8\s2\t2 +\s2\u2\nonumber\\
&&      -4\s2-\t2^2\u2+4\t2^2+2\t2\u2-8\t2-\u2+4))
\end{eqnarray}
The apropriate colour factors are:
\begin{eqnarray}
&&C_{11}=C_{22}=C_{12}=C_{21}=64/9,\nonumber\\
&&C_{33}=C_{44}=C_{34}=C_{43}=1/9,\nonumber\\
&&C_{13}=C_{31}=C_{14}=C_{41}=C_{23}=C_{32}=C_{24}=C_{42}=-8/9,
  \nonumber\\
&&C_{15}=C_{51}=C_{25}=C_{52}=8,\nonumber\\
&&C_{35}=C_{53}=C_{45}=C_{54}=-1, \qquad  C_{55}=-9
\end{eqnarray}

The differential cross section for subprocess $g c\to J/\psi c$
can be written as follows:
\begin{equation}
 \frac{d\hat\sigma}{d\hat t}=\frac{|\bar M|^2}{16\pi(\hat s-m^2)^2}
  \end{equation}

In the general factorization approach of QCD the measurable cross
section
$\sigma(p p\to J/\psi c X)$ and the partonic cross section
$\hat\sigma(g c\to J/\psi c)$ are connected by the following expression:
\begin{eqnarray}
\frac{d\sigma}{d^2p_{\perp}dy}&=&\int dx_1\int dx_2 \left[G_p(x_1,Q^2)
 C_p(x_2,Q^2)+(1\Leftrightarrow 2)\right]\nonumber\\
 &&\frac{x_1x_2s}{\pi}
 \delta(\hat s+\hat t+\hat u-6m^2)\frac{d\hat\sigma}{d\hat t}.
\end{eqnarray}

Here:
 $$ \hat t=M_J^2+m^2-x_2(M_J^2-t),$$
$$ \hat u=M_J^2-x_1(M_J^2-u),$$
$$\hat s=x_1x_2s+m^2,~~~~~ M_\perp=\sqrt{M_J^2+p_\perp^2},~~~~~~
   M_J=2m,$$
$$t=M_J^2-\sqrt{s}M_\perp \exp(y),$$
$$u=M_J^2-\sqrt{s}M_\perp \exp(-y),$$
where $y$ is $J/\psi$ rapidity in s.c.m., $p_\perp$ is $J/\psi$
transverse momentum, $G_p(x,Q^2)$ and $C_p(x,Q^2)$ are gluon and
$c$-quark
distribution function in a proton (antiproton) at the scale
$Q^2=M_\perp^2$.


For comparison we plot in Fig.2 the partonic cross sections for
subprocesses
$g c\to J/\psi c$ and $g g\to J/\psi g$ versus value of the scaled
 momentum
 transfer, $-\hat t/\hat s$, for typical value of the subenergy,
 $\hat s=50$
 GeV$^2$. We observe from Fig.2 that cross section of c-quark--gluon
 scattering
 subprocess is larger than gluon--gluon cross section on factor 2-50
in the wide region of $-\hat t/\hat s$. The comparebly large cross
section of $cg\to J/\psi c$ subprocess compensates in part the small
value of the charm quark structure function in the proton.

Fig.3 shows the result of our calculation for $J/\psi$ with large
transverce momentum ($p_T>5$ GeV/c) plus open charm
associated production in $pp(p\bar p)$ interaction versus $\sqrt s$.
We use in calculations "hard" (nonperturbative) scaling parameterization
\cite{13a} and two set of "soft" (perturbative) parameterization from
 refs.
\cite{13}(DO) and \cite{13b}(GRV)  for $C_p(x,Q^2)$ structure functions
 and take the usual value   of QCD $K-$factor:  $K=2$. Note that the
 mean value of the proton momentum, which is carried by charm quarks,
is approximately equal for all parameterization (0.3-0.5\%), and does
not  contradict data from EMC collaboration on $F_2^c(x,Q^2)$ \cite{14}.


We observe from Fig.3 that the contribution for "hard" parameterization
\cite{13a} dominants at small energy region $\sqrt s<70$ GeV, opposite
 at
 more high energies the contribution for "soft" parameterization
 \cite{13}
is large and it grows from .1 nbn at $\sqrt s=100$ GeV to 4 nbn at
$\sqrt s=1$ TeV.
 Such a way we predict sufficiently large measurable cross section for
 $J/\psi$ plus open charm production at hadron colliders via partonic
  subprocess $g c\to J/\psi c$. It gives us the
  opportunity of experimental investigation of the x-dependence of charm
  quark distribution function in the proton, especcialy at small-x
  region.

%

Figs. 4 and 5 show our predictions for $J/\psi 's$ $p_T$ and $y$-spectra
at the energy $\sqrt s=630$ GeV in $p\bar p$ interactions. Curver 1 in
Figs.4 and 5 is the result of calculation for intrinsic charm
contribution,
curver 2 is the contribution for direct quarkonium production via
gluon-gluon
fusion and curver 3 is the contribution for $J/\psi$ from decay
$B\to J/\psi X$.
Curvers 2 and 3 are taken from ref. \cite{2}.

{\it\bf Acknowledgements.} This  research was supported by the Russian
  Foundation of Basic Research (Grant 93-02-3545)
  and by State Committee on High Education of Russian Federation
  (Grant 94-6.7-2015). Authors thank S.~Baranov,
 N.~Zotov and A.~Likhoded  for useful discussions.

{\large\bf Figure captions.}

\begin{enumerate}
\item Diagrams used to describe the partonic subprocess $gc\to
 J/\psi c$.
\item The partonic cross sections as function of $x=-\hat t/\hat s$
at $\hat s=50$ GeV$^2$. The  curver 1 is $gc\to J/\psi c$ subprocess,
 the curver 2 is $gg\to J/\psi g$ subprocess.
\item The total cross section for $J/\psi$ plus open charm associated
 production via intrinsic charm in $pp$ or $p\bar p$ collisions.
 The curvers 1-3 correspond to parameterizations charm quark structure
  functions from refs.\cite{13a,13,13b}
\item The $p_\perp$ distribution for $J/\psi$ production in $p\bar p$
 collisions at $\sqrt s=630$ GeV and $|y|<2$. The data are from \cite{5} .
 The curver 1 is intrinsic charm quark contribution.
 The curver 2 is the sum of the direct charmonium production
 subprocesses.
 The curver 3 corresponds to $J/\psi$'s of $b\bar b$ origin.
 The curver 4 is the sum of all contributions.
\item
The $y$ distribution for $J/\psi$ production in $p\bar p$
 collisions at $\sqrt s=630$ GeV and $p_\perp>5$ GeV. The  same as in
 Fig.~4.
\end{enumerate}

\begin{thebibliography}{99}
\bibitem{1} R.~Baier, R.~Ruckl. Nucl.Phys.B218~(1983)~289; B20~(1982)~1.
\bibitem{2} E.W.N.~Glover, A.D.~Martin, W.J.~Stirling. Z.Phys.C38~(1988)
 ~473.
\bibitem{3} R.~Vogt, S.J.~Brodsky, P.~Hoyer. Preprint SLAC-Pub-5827,
 May, 1992.\\
 G.A.~Schuler. Preprint CERN-TH.7170/94 (1994)\\
 V.~Barger, W.Y.~Keung, R.J.N.~Phyllips.Phys.Let.91B~(1980)~253,\\
 K.~Harigiwara, A.D.~Martin, W.J.~Stirling. Phys.Lett.B267~(1991)~527.
\bibitem{4} A.G.~Clark et al. Nucl.Phys.B142~(1978)~29,\\
 C.~Kourkoumelis et al. Phys.Lett.91B~(1980)~481.
\bibitem{5} Albajar et al. UA1 Collab. Preprint CERN-EP/87-175, 1987.
\bibitem{6} F.~Abe et al. CDF Collab. Preprint Fermilab -Pub-92/236-E,
 1992.
\bibitem{7} F.~Halzen et al. Phys.Rev.D30~(1984)~700,\\
 R.~Gastmans, W.~Troost, T.T.~Wu. Phys.Lett.B184~(1987)~257.
\bibitem{8} E.~Braaten, T.C.~Yuan. Phys.Rev.Lett.71~(1993)~1673,\\
M.A.~Doncheski, S.~Fleming, M.L.~Mangano. Preprint Fermilab-Conf-93/
348-T,
 1993.
\bibitem{9} E.L.~Berger. Z.Phys.C4~(1980)~289.
\bibitem{10} V.A.~Saleev. In Proc. of Conf. "Non-commutative structures
  in mathematical physics", Togliatti, October, 1993; Mod.Phys.Lett.
  A9~(1994)~1083.

 \bibitem{11} V.~Barger, F.~Halzen, W.Y.~Keung. D25~(1982)~112;\\
  E.Hoffmann, R.Moore. Z.Phys.C20~(1983)~71;\\
  G.Ingelman, L.Jonsson, M.Nyberg. Preprint DESY 92-178, 1992.
  \bibitem{12} Rewiew of Particle Properties. Phys.Rev.D45~(1992).
  \bibitem{13a} S.J.Brodsky et al. Phys.Lett.B93~(1980)~451;
   S.J.Brodsky,  C.Peterson. Phys.Rev.D23~(1981)~2745.
 \bibitem{13} D.W.~Duke, J.F.~Owens. Phys.Rev.D30~(1984)~49.
 \bibitem{13b} M.Gluck, E.Reya, A.Vogt. Phys.Rev. D45~(1992)~3986;
 D46~(1992)~1973; A.Vogt. Preprint DO-TH 92/15, Dortmund, 1992.
 \bibitem{14} Aubert~J.J. et al. Phys.Lett.B110~(1982)~73;
\end{thebibliography}
\end{document}